\documentclass{pasj01}
\Received{$\langle$reception date$\rangle$}
\Accepted{$\langle$acception date$\rangle$}
\Published{$\langle$publication date$\rangle$}

\begin{document}

\title{
A search for Ly$\alpha$ emitters around a concentrated region of strong Ly$\alpha$ absorbers at $z=2.3$
\thanks{Based on data collected at the Subaru Telescope, which is operated by the
National Astronomical Observatory of Japan.}
}

\author{Kazuyuki \textsc{Ogura}\altaffilmark{1}}
\author{Tohru \textsc{Nagao}\altaffilmark{2}}
\author{Masatoshi \textsc{Imanishi}\altaffilmark{3, 4, 5}}
\author{Nobunari \textsc{Kashikawa}\altaffilmark{4, 5}}
\author{Yoshiaki \textsc{Taniguchi}\altaffilmark{6}}
\author{Masaru \textsc{Kajisawa}\altaffilmark{1, 2}}
\author{Masakazu A. R. \textsc{Kobayashi}\altaffilmark{7}}
\author{Yoshiki \textsc{Toba}\altaffilmark{2, 8}}
\author{Kodai \textsc{Nobuhara}\altaffilmark{1}}

\altaffiltext{1}{Graduate School of Science and Engineering, Ehime University, 
 2-5 Bunkyo-cho, Matsuyama, Ehime 790-8577, Japan }
\altaffiltext{2}{Research Center for Space and Cosmic Evolution, Ehime University, 
 2-5 Bunkyo-cho, Matsuyama, Ehime 790-8577, Japan }
\altaffiltext{3}{Subaru Telescope, 650 North A'ohoku Place, Hilo, HI 96720, USA }
\altaffiltext{4}{National Astronomical Observatory of Japan, 2-21-1 Osawa, Mitaka, Tokyo 181-8588, Japan }
\altaffiltext{5}{Department of Astronomy, School of Science, SOKENDAI (The Graduate University for Advanced
Studies), 2-21-1 Osawa, Mitaka, Tokyo 181-8588, Japan }
\altaffiltext{6}{The Open University of Japan, 2-11, Wakaba, Mihama-ku, Chiba, Chiba 261-8586, Japan }
\altaffiltext{7}{Faculty of Natural Sciences, National Institute of Technology, Kure College, 2-2-11 Agaminami, Kure, Hiroshima 737-8506, Japan }
\altaffiltext{8}{Academia Sinica Institute of Astronomy and Astrophysics, PO Box 23-141, Taipei 10617, Taiwan }

\email{ogura@cosmos.phys.sci.ehime-u.ac.jp}


\maketitle

\begin{abstract}
In order to investigate the physical relationship between strong Ly$\alpha$ absorbers (log$N_{\rm HI} \geq 20.0$~cm$^{-2}$)
such as damped Ly$\alpha$ absorption 
systems (DLAs) and young star-forming galaxies at high redshift, we have conducted narrow-band observations of Ly$\alpha$ 
emitters (LAEs) in a concentrated region of strong Ly$\alpha$ absorbers at $z=2.3$, the J1230+34 field.
Using a catalog of Ly$\alpha$ absorbers with log$N_{\rm HI} \geq 20.0$~cm$^{-2}$ based on the baryon oscillation spectroscopic 
survey (BOSS), we found 6 fields where 3 or more absorbers are concentrated within a (50~Mpc)$^3$ cubic box in the comoving scale. 
Among them, we focus on the J1230+34 field, where 2 DLAs and 2 sub-DLAs present. 
Our narrow-band imaging observations with Subaru/Suprime-Cam using a custom-made filter, $NB400$ ($\lambda_{\rm c} =4003$~\AA~and 
FWHM$=92$~\AA) yield a sample of 149 LAEs in this field.
In the large scale ($\sim$50 Mpc), we have found no differences between the obtained Ly$\alpha$ luminosity function and those in the 
blank fields at similar redshifts. We also compare the frequency distribution of the Ly$\alpha$ rest-frame equivalent width ($EW_{0}$) 
in the target field and other fields including both overdensity region and blank field, but find no differences.
On the other hand, in the small scale ($\sim$10 Mpc), 
we have found a possible overdensity of LAEs around a DLA with the highest H~{\sc i} column
density ($N_{\rm HI} = 21.08$~cm$^{-2}$) in the target field while there are no density excess around the other absorbers with a 
lower $N_{\rm HI}$.
\end{abstract}

\section{Introduction}
Investigating the formation and evolution of galaxies is one of the important topics in the modern 
astrophysics. Especially, it is interesting to study how cold gas is converted to stars
in galaxies at the very early phase of evolution.
To investigate the very early phase of the galaxy evolution, we focus on two populations that are thought
to be gas-rich young systems, the damped Ly$\alpha$ absorption system (DLA) and the Ly$\alpha$ emitter (LAE). 

The DLA is a class of quasar absorption-line systems, which has a column density of 
$N_{\rm HI} \geq 10^{20.3}$ cm$^{-2}$ (Wolfe et al. 2005). 
The DLA provides a powerful tool to investigate the nature of the cold gas at high-$z$
because they trace the intervening gas along the line-of-sight to quasars and can be detected as a strong 
Ly$\alpha$ absorption line on quasar spectra regardless of the luminosity of their stellar component.
Since DLAs dominate the neutral-gas content in a wide redshift range (Storrie-Lombardi \& Wolfe 2005), 
they are thought to be gas reservoirs for the star-formation in the high-$z$ Universe.
Though the DLA is such an important population, their nature is still under the debate 
(e.g., Prochaska \& Wolfe 1997; Taniguchi \& Shioya 2000; 2001; Kacprzak et al. 2010)
because it is often difficult to identify their optical counterparts due to their faintness, especially at high redshift 
(see for recent identifications at $z > 2$, 
Fynbo et al. 2010; 2011; 2013; P{\'e}roux et al. 2011; 2012; Bouch{\'e} et al. 2013;
Krogager et al. 2012, 2013, Kashikawa et al. 2014, Srianand et al. 2016; Sommer-Larsen \& Fynbo 2017 and references therein).
In many cases, counterparts of high-$z$ DLAs have been found as LAEs. This is consistent with the idea that the nature of DLAs is
young gas-rich galaxy.

The LAE is a population of galaxies selected by their strong Ly$\alpha$ emission. 
Since typical LAEs show faint UV continuum and large Ly$\alpha$ equivalent width ($EW$), it is considered that they are young 
galaxies (e.g., Malhotra \& Rhoads 2002, Taniguchi et al. 2005; Shimasaku et al. 2005; Kashikawa et al. 2006; Gawiser et al. 2007; 
Nilsson et al. 2007; Ono et al. 2010b; 2012).
So far, various researches to understand properties of LAEs have been conducted and found that their typical stellar mass 
is $10^8 - 10^9~M_{\odot}$ and the age is in order of 100~Myr (e.g., Gawiser et al. 2006; 2007; Nilsson et al. 2009; Ono et al. 2012).
Note that recently, some LAEs with more evolved stellar populations have been found (e.g., Ono et al. 2010a; Taniguchi et al. 2015).
In addition, some sensitive radio observations of CO molecular lines in high-$z$ galaxies suggest that the gas fraction is larger at higher redshifts, 
and sometimes the gas mass fraction reaches up to $\sim$0.5 or even more (e.g., Carilli \& Walter 2013; Troncoso et al. 2014).
Therefore the LAE as well as the DLA at high $z$ is a key population to understand the early phase of the galaxy evolution.
However, the relation between these two populations is still unclear.
As we mention above, although most of DLA counterparts at high $z$ show the Ly$\alpha$ emission, only few counterparts of
high-$z$ DLAs have been identified so far.
Given the idea that DLAs are neutral gas reservoir for the star formation, the relation of the DLAs and young galaxies (i.e., LAEs) is interesting. 

It should be noted that DLAs can be recognized only when they have a background quasar (BGQSO). 
The number density of such quasars had been too low to investigate grouping or clustering properties of DLAs.
However, such a situation has been recently changed after the data release of Baryon Oscillation 
Spectroscopic Survey (BOSS; Eisenstein et al. 2011; Dawson et al. 2013). The quasar catalog based on 
the BOSS data (Data Release 9 of the Sloan Digital Sky Survey [SDSS DR9]; Ahn et al. 2012) includes 
87,822 quasars mainly at $2 < z < 4$ (P{\^ a}ris et al. 2012).
By utilizing the BOSS quasar sample, Noterdaeme et al. (2012) found 12,801 absorbers with $N_{\rm H~I} > 10^{20.0} {\rm cm}^{-2}$
(partly including sub-DLAs which are absorption-line systems with $10^{19.0} {\rm cm}^{-2} < N_{\rm H~I} < 10^{20.3} {\rm cm}^{-2}$;
P{\'e}roux et al. 2003). 
This new and large absorber sample enables us to study the spatial distribution of DLAs.

Here, we focus on regions where some absorbers with a narrow redshift range are concentrated in a narrow area.
Such a region is interesting for following two reasons.
(1) Such a region may harbor large amount of neutral gas and thus it may correspond to the region where
galaxies are rapidly evolving, and (2) to observe such a region, we can effectively search for counterparts of absorbers.  
By observing LAEs around such a region, we can investigate the relation of gas-rich systems and young galaxies.
For this purpose, wide-field narrow-band (NB) observations are very useful.
By combining NB filters and imaging instruments with a wide field-of-view, we can effectively observe LAEs in a wide area.

In this paper, we report our observational trial for studying properties of LAEs around a concentrated region of strong
Ly$\alpha$ absorbers to investigate very early stage of the galaxy evolution based on the NB imaging observations.
This paper is organized as follows.
In Section 2, we present the target field selection and imaging observations of LAEs.
The results are given in Section 3. We discuss interpretations of the absorber concentrated region in Section 4.
We present our conclusion in Section 5.
We use a $\Lambda$ CDM cosmology with $H_{0} = 70~{\rm km~s}^{-1}$~Mpc$^{-1}$, 
$\Omega_{\rm M} = 0.3$, and $\Omega_{\rm \Lambda} = 0.7$, throughout this paper. 
Magnitudes are all given in the AB system.

\section{The target field and observations}
\subsection{The target field}
\begin{figure}
 \begin{center}
  \includegraphics[width=9cm]{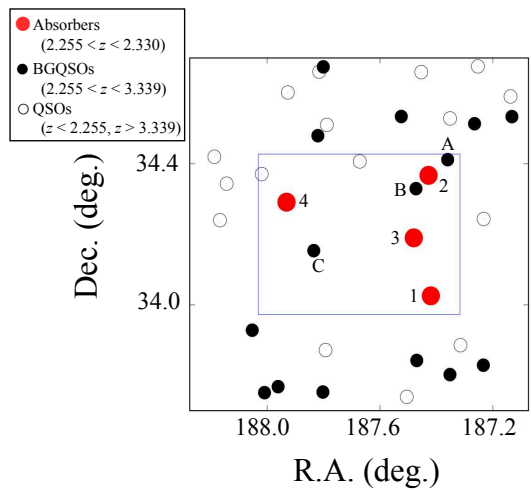} 
 \end{center}
\caption{The sky distribution of absorbers in the J1230+34 field ($1^{\circ} \times 1^{\circ}$). 
The blue box shows the field-of-view of Suprime-Cam ($34^{\prime} \times 27^{\prime}$) in our observation. 
Filled circles show line-of-sights (LoSs) to background quasars (BGQSOs).
Red and black filled circles show those with absorber and without absorbers at $2.255 < z < 2.330$.
IDs (1-4) for the absorbers (filled red circles) are same as in Table \ref{tab:info_j1230}.
We label BGQSOs without absorber at $2.255 < z < 2.330$ as A to C.
Open circles show LoSs to BOSS quasars that do not satisfy the criteria for BGQSOs ($z<2.255$ or $z>3.339$).
%
}\label{fig:map_j1230}
\end{figure}

\begin{figure*}
 \begin{center}
  \includegraphics[width=12cm]{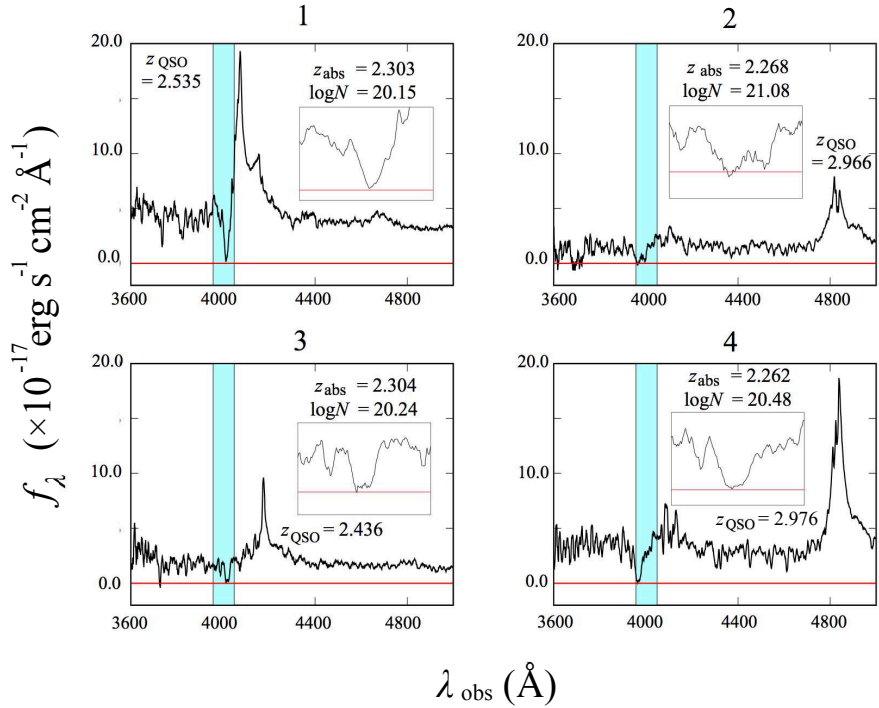} 
 \end{center}
\caption{BOSS spectrum of quasars showing a strong intervening Ly$\alpha$ absorption line at $2.255 < z < 2.330$ 
in the J1230+34 field. Cyan shadowed region in each panel shows the wavelength coverage of $NB400$.
Horizontal red lines correspond to $f_{\lambda}=0$.
Inserted panels show the magnified view around absorbing feature.
}\label{fig:spDLA}
\end{figure*}

In this study, we focus on the J1230+34 field, where there are 2 DLAs and 2 sub-DLAs at $z\sim2.3$ within a cubic region
of $(\sim50 {\rm Mpc})^3$. Table \ref{tab:info_j1230} shows the basic data of four absorbers, and Figure \ref{fig:map_j1230} 
shows the sky distribution of the absorbers and quasars in the J1230+34 field. The BOSS spectra of quasars with a strong absorption in the 
target field is shown in Figure \ref{fig:spDLA}.
We selected this region from the catalog of strong ($N_{\rm H~I} > 10^{20.0} {\rm cm}^{-2}$) Ly$\alpha$ absorbers
based on the BOSS quasars (Noterdaeme et al. 2012).

\begin{figure}
 \begin{center}
  \includegraphics[width=8cm]{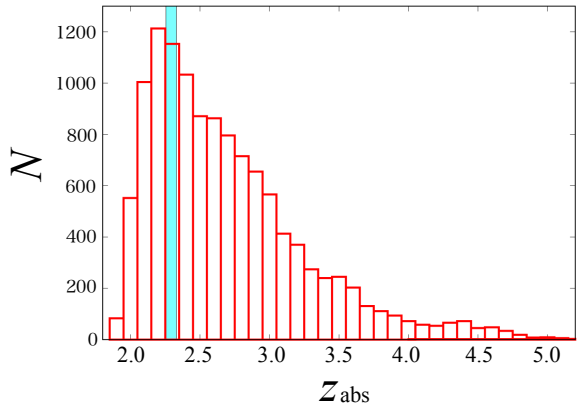} 
 \end{center}
\caption{The redshift distribution of absorbers with $N_{\rm HI} > 10^{20.0}$ cm$^{-2}$ in the catalog based on the 
BOSS (Noterdaeme et al. 2012). Cyan shadow region shows the redshift coverage of $NB400$ ($2.255 < z < 2.330$).
}\label{fig:zDLA}
\end{figure}

Here, we describe the selection process of the target field.
First, we define the sample of absorbers and their BGQSOs.
The redshift distribution of the absorbers in the catalog shows its peak at $z \sim$2.2 (Figure \ref{fig:zDLA}). 
Fortunately, there is a suitable narrow-band (NB) filter of Suprime-Cam (Miyazaki et al. 2002) on the 8.2 m Subaru 
telescope, $NB400$, to observe the Ly$\alpha$ emission from objects around that redshift. 
$NB400$ is a custom-made NB filter whose central wavelength and width in full width at half maximum 
(FWHM) are 4003~\AA~and 92~\AA, respectively. This wavelength coverage corresponds to the redshift 
range of $2.255 < z < 2.330$ for the Ly$\alpha$ emission (1216~\AA~in the rest frame). 
Therefore, throughout this paper, we define the absorber sample as those at $2.255<z<2.330$.
As for the BGQSOs, we set their redshift range to $2.255 < z < 3.339$. The lower limit of the redshift range for BGQSOs 
is the same as that for absorbers, taking also into account of proximity DLAs (PDLAs; Prochaska et al. 2008). 
The PDLA is a population of DLAs at a velocity separation of $<$ 3,000 km~s$^{-1}$ from the BGQSO (Prochaska et al. 2008). 
We determine the upper limit of the redshift of BGQSOs by considering the Lyman limit of BGQSOs. Since the signal-to-noise (S/N) 
ratio of a quasar spectrum is very low at the wavelength range shorter than the Lyman limit (912~\AA~in the rest frame), 
it is difficult to find absorbers at this wavelength range. Therefore we focus only on quasars at the redshift where their Lyman limit 
locates at shorter wavelength than the $NB400$ coverage. We remove quasars with the balnicity index (BI; Weymann et al. 1991) 
$>$1,000 km~s$^{-1}$ from the BGQSO sample, because it is hard to distinguish Ly$\alpha$ absorbers from broad absorption 
lines (BALs) in the intrinsic spectra of such BAL quasars (Noterdaeme et al. 2012).
Among the BOSS quasar sample (87,822 quasars at the time of DR9), we regard 47,376 quasars as BGQSOs for absorbers at 
$2.255 < z < 2.330$. The number of absorbers at $2.255 < z < 2.330$ whose BGQSOs satisfy the criteria is 824.

Then we search for concentrated regions of strong Ly$\alpha$ absorbers based on this sample.
Here, we define the concentrated region as the region where there are three or more absorbers within a cubic space of 
$(50~{\rm Mpc})^3$.
This definition is based on the typical size of galaxy proroclusters. Overzier (2016) shows the redshift distribution of galaxies
in several protoclusters which are confirmed spectroscopically. The width of redshift distribution of galaxies in protoclusters at $z\sim2$ 
is typically $\Delta z\sim 0.02-0.06$ (corresponding to $\sim26-76$ Mpc in the comoving scale). Chiang et al. (2013) 
also summarized properties of observed protoclusters and most of them show $\Delta z \sim 0.03-0.06$ (corresponding to 
$\sim38-76$ Mpc in the comoving scale). 
As for the spatial extent of protoclusters, it is known that they could be extended up to dozens of Mpc (e.g., Hayashino et al. 2004; 
Prescott et al. 2008; Lee et al. 2014). Recently, Franck and McGaugh (2016) search for clusters and protoclusters over a wide redshift 
range ($2.7 < z < 3.7$), showing that almost all members of a cluster are covered with the circular region whose radius is 
$\sim$20~Mpc in the comoving scale.
In addition, Muldrew et al. (2015) studied to define the high-$z$ protoclusters and found that 90\% of a protocluster mass
is extended across 50 comoving Mpc at $z\sim2$. 
Taking considerations given above into account, we decide the search scale of (50~Mpc)$^3$ in this work, 
and define the concentrated regions of strong Ly$\alpha$ absorbers as the region with 3 or more absorbers within (50~Mpc)$^3$.

\begin{figure*}
 \begin{center}
  \includegraphics[width=15cm]{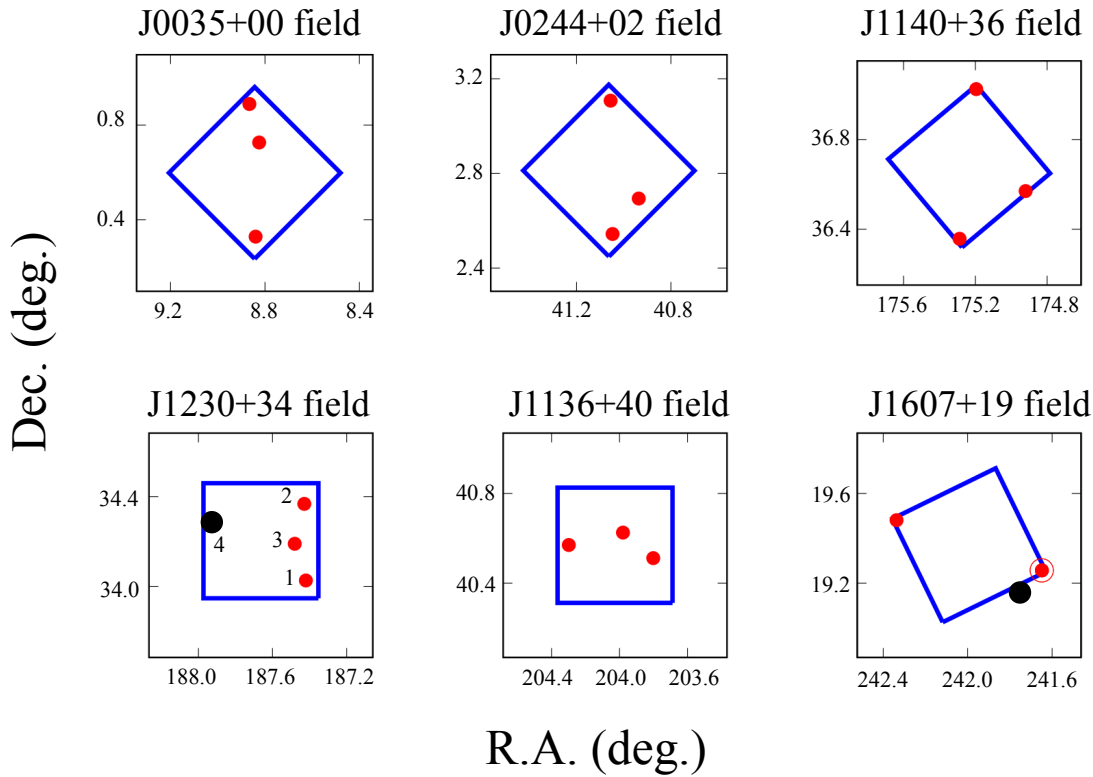} 
 \end{center}
\caption{The sky distribution of absorbers in each absorber-concentrated region. 
Red filled circles show absorbers. Blue boxes show square regions with 50 Mpc (in the comoving scale at $z=2.3$) on a side.
The red filled circle with open red circle in the J1607+19 field denote the quasar sight line with 2 absorbers within the concentrated region.
Black filled circles in the J1230+34 field and the J1607+19 field show absorbers at $2.255<z<2.330$ out of the (50~Mpc)$^{3}$ region.
IDs for absorbers in the J1230+34 field are same as in Table \ref{tab:info_j1230} and Figure \ref{fig:map_j1230}.
}\label{fig:all_map}
\end{figure*}

\begin{figure}
 \begin{center}
  \includegraphics[width=8.5cm]{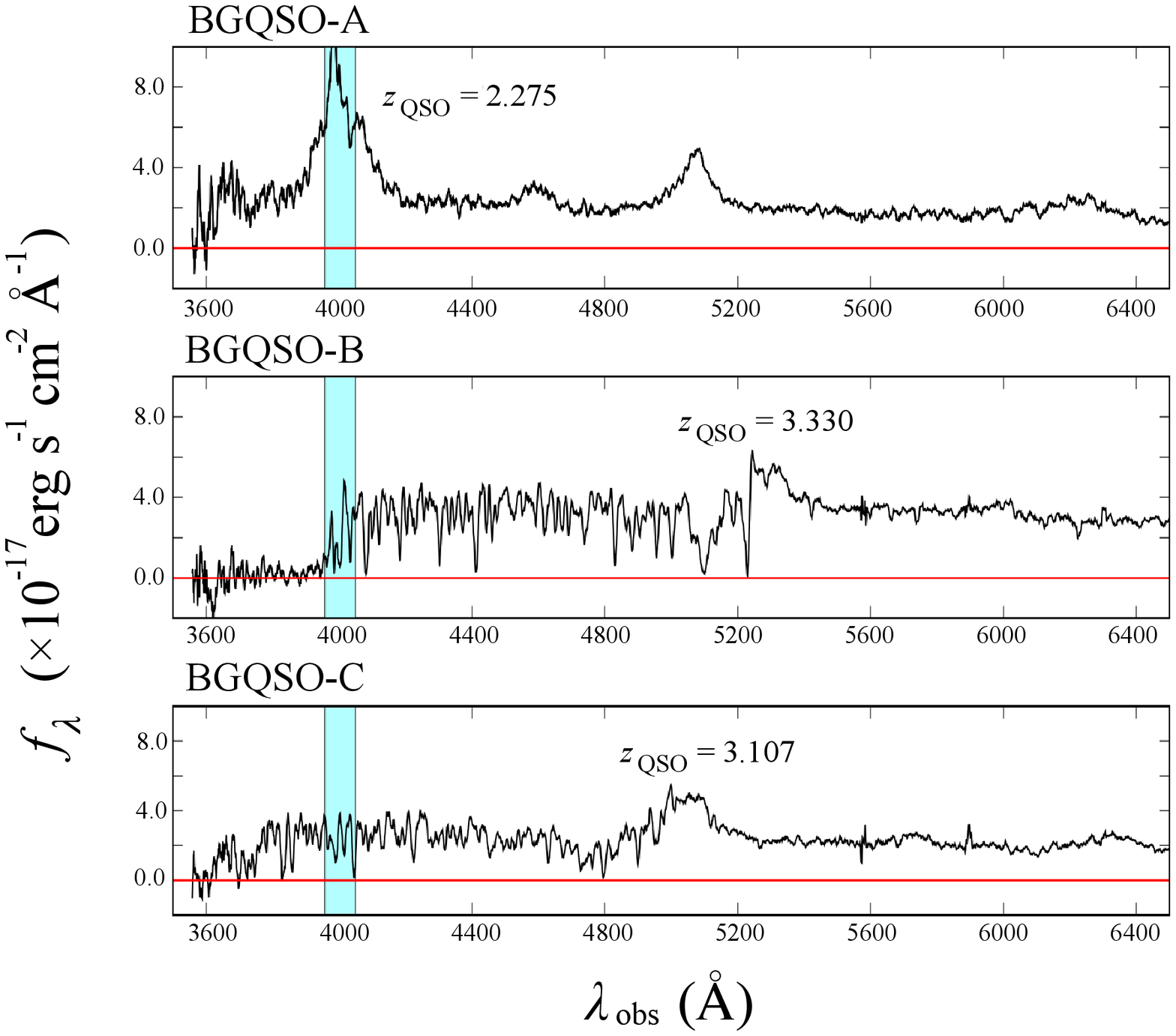} 
 \end{center}
\caption{BOSS spectrum of quasars not showing strong intervening Ly$\alpha$ absorption-line at $2.255 < z < 2.330$
in the J1230+34 field. Cyan shadowed regions in each panel show the wavelength coverage of $NB400$.
Horizontal red lines correspond to $f_{\lambda}=0$. Although BGQSO-A at $z=2.275$ is a member of objects at $z\sim2.3$,
it satisfies the criteria for BGQSO since we determine the criteria taking PDLAs into account. Note that, BGQSO-A
is detected as a LAE in our observation (see Section 3.3.3).  
}\label{fig:spBGQSO}
\end{figure}

As a result, we have found 6 absorber-concentrated regions in which there are at least 3 strong Ly$\alpha$ absorbers.
Table \ref{tab:concentration} shows the basic data of absorbers in each region and Figure \ref{fig:all_map} shows
the sky distribution of absorbers.
As shown in Table \ref{tab:concentration} (and also Figure \ref{fig:all_map}), all of selected fields have 3 absorbers.
Interestingly, there is one more absorber which is located at very close to 3 absorbers in the J1230+34 field.
This absorber (No. 4 in Table \ref{tab:info_j1230}, Figure \ref{fig:map_j1230} and \ref{fig:all_map}) is located
within the 50 Mpc $\times$  50 Mpc box on the sky. 
Furthermore, the comoving distance along the line of sight from absorber No. 4 to the nearest absorber (No. 2) is only 7.7 Mpc and that
between No. 4 and farthest absorber (No. 1) is 53.5 Mpc.
If considering also this absorber, there are more absorbers in the J1230+34 field than the other 5 fields, and thus this J1230+34 
field is more interesting as an absorber-concentrated region than the other fields.
The rarity of this field is shown in Section 2.2.
Seven quasars in this region satisfy the criteria for BGQSOs, and 3 (A, B, and C in Figure \ref{fig:map_j1230}) among them 
do not have absorbers at $2.255<z<2.330$. Figure \ref{fig:spBGQSO} shows BOSS spectrum of these BGQSOs.
We describe our NB observations of this field in Section 2.4.


\begin{longtable}{cccccccc}
  \caption{The information of absorbers in the J1230+34 field}
  \label{tab:info_j1230}
  \hline
  No. & Quasar Name & R.A. (deg.) & Dec. (deg.) & $z_{\rm QSO}$ & $z_{\rm DLA}$ & log$N_{\rm HI}$ (cm$^{-2}$) & Class \\ 
  \endfirsthead
  \hline
  No. & Quasar Name & R.A. (deg.) & Dec. (deg.) & $z_{\rm QSO}$ & $z_{\rm DLA}$ & log$N_{\rm HI}$ (cm$^{-2}$) & Class \\
  \endhead
  \hline
  \endfoot
  \hline
 \multicolumn{8}{l}{{\bf Notes.}~Errors on $z_{\rm QSO}$, $z_{\rm DLA}$ and $N_{\rm HI}$ are not provided by Noterdaeme et al. (2012).}\\
  \endlastfoot
  \hline
 1 & SDSS J122940.83+340132.3 & 187.4201 & 34.0256 & 2.353 & 2.303 & 20.15 & sub-DLA\\
 2 & SDSS J122942.74+342202.1 & 187.4281 & 34.3673 & 2.966 & 2.268 & 21.08 & DLA \\
 3 & SDSS J122955.14+341123.7 & 187.4798 & 34.1899 & 2.436 & 2.304 & 20.24 & sub-DLA\\
 4 & SDSS J123143.46+341727.6 & 187.9311 & 34.2910 & 2.976 & 2.262 & 20.48 & DLA \\
 \end{longtable} 

 \begin{longtable}{c|ccccc}
  \caption{Basic data of the absorber-concentrated regions}
  \label{tab:concentration}
  \hline
  Field & R.A. (deg.) & Dec. (deg.) & $z_{\rm QSO}$ & $z_{\rm abs}$ & log$N_{\rm HI}$ (cm$^{-2}$) \\
  \endfirsthead
    \hline
  Field & R.A. (deg.) & Dec. (deg.) & $z_{\rm QSO}$ & $z_{\rm abs}$ & log$N_{\rm HI}$ (cm$^{-2}$) \\
  \endhead
  \hline
  \endfoot
  \hline
   \multicolumn{6}{l}{{\bf Notes.}~Errors on $z_{\rm QSO}$, $z_{\rm DLA}$ and $N_{\rm HI}$ are not provided by Noterdaeme et al. (2012).}\\
  \endlastfoot
  \hline
J0035+00 field & 8.8248 & 0.7260 & 2.925 & 2.304 & 20.40  \\
& 8.8400 & 0.3280 & 2.360 & 2.311 & 20.70 \\
& 8.8653 & 0.8892 & 2.460 & 2.282 & 20.14 \\
 \hline
J0244+02 field & 40.9367 & 2.6943 & 2.806 & 2.264 & 20.16 \\
& 41.0469 & 2.5439 & 2.668 & 2.257 & 20.78 \\
& 41.0558 & 3.1079 & 2.289 & 2.267 & 20.07 \\
 \hline
J1140+36 field & 174.9200 & 36.5705 & 2.904 & 2.279 & 20.56 \\
& 175.2880 & 36.3572 & 2.274 & 2.271 & 20.07 \\
& 175.1960 & 37.0252 & 2.837 & 2.303 & 20.15 \\
 \hline
J1230+34 field & 187.4201 & 34.0256 & 2.353 & 2.303 & 20.15 \\
& 187.4281 & 34.3673 & 2.966 & 2.268 & 21.08 \\
& 187.4798 & 34.1899 & 2.436 & 2.304 & 20.24 \\
 \hline
J1336+40 field & 203.8010 & 40.5116 & 2.832 & 2.268 & 20.63 \\
& 204.2980 & 40.5707 & 2.655 & 2.291 & 20.03 \\
& 203.9800 & 40.6252 & 2.288 & 2.266 & 20.11 \\
 \hline
J1606+19 field & 241.6490 & 19.2572 & 2.528 & 2.282 & 20.00 \\
& 241.6490 & 19.2572 & 2.528 & 2.320 & 21.40 \\
& 242.3370 & 19.4813 & 2.844 & 2.297 & 21.40 \\
\end{longtable} 

\subsection{The rarity of the target field}
Here, we show how rare the target region (the J1230+34 field) is.
As we described in Section 2.1, we define the concentrated region as a region with 3 or more absorbers
within the cubic space of (50 Mpc)$^3$ in the comoving scale.
To investigate the rarity of the target field, we examine how many fields are selected for different volumes.
In Table \ref{tab:N_field}, we show numbers of selected field when we apply various criteria.
Regions harboring 4 or more absorbers are extremely rare. The J1230+34 field is the only field harboring 4 absorbers even 
if we adopt a larger size of (60 Mpc)$^3$, and only 2 regions are selected even in the case of (70 Mpc)$^3$.
Based on this, we conclude that the J1230+34 field is an extremely rare and interesting region.

\begin{table}
  \tbl{Numbers of absorber concentrated regions in various volumes}{%
  \begin{tabular}{llll}
      \hline
      Volume (Mpc$^3$) & $N_{\rm abs}\geq3$\footnotemark[a] & $N_{\rm abs}\geq4$\footnotemark[b]\\ 
      \hline
      30$^3$ & 0 & 0 \\ 
      40$^3$ & 3 & 0 \\ 
      50$^3$ & 6 & 0 \\ 
      60$^3$ & 15 & 1 \\ 
      70$^3$ & 27 & 2 \\ 
       \hline
    \end{tabular}}\label{tab:N_field}
\begin{tabnote}
{\bf Notes.} \\
\footnotemark[a] The number of the concentrated region with 3 or more absorbers. \\
\footnotemark[b] The number of the concentrated region with 4 ore more absorbers. Note that there are no fields with 5 or more absorbers.
\end{tabnote}
\end{table}

\subsection{The surface number density distribution of absorbers at $2.255<z<2.330$ and BGQSOs}

Since a quasar absorption-line system can be found only when it has a BGQSO, the sky distribution of BGQSOs is 
particularly important to discuss that of absorbers. In a region where the surface number density of 
BGQSOs is relatively high, it is expected that the apparent number of observable absorbers is accordingly higher than regions 
with a lower surface number density of  BGQSOs. Here, we investigate the surface number densities of both absorbers 
and BGQSOs in each absorber concentrated region to examine whether or not the spatial distribution of BGQSOs influences that of absorbers.

\begin{figure}
 \begin{center}
  \includegraphics[width=8cm]{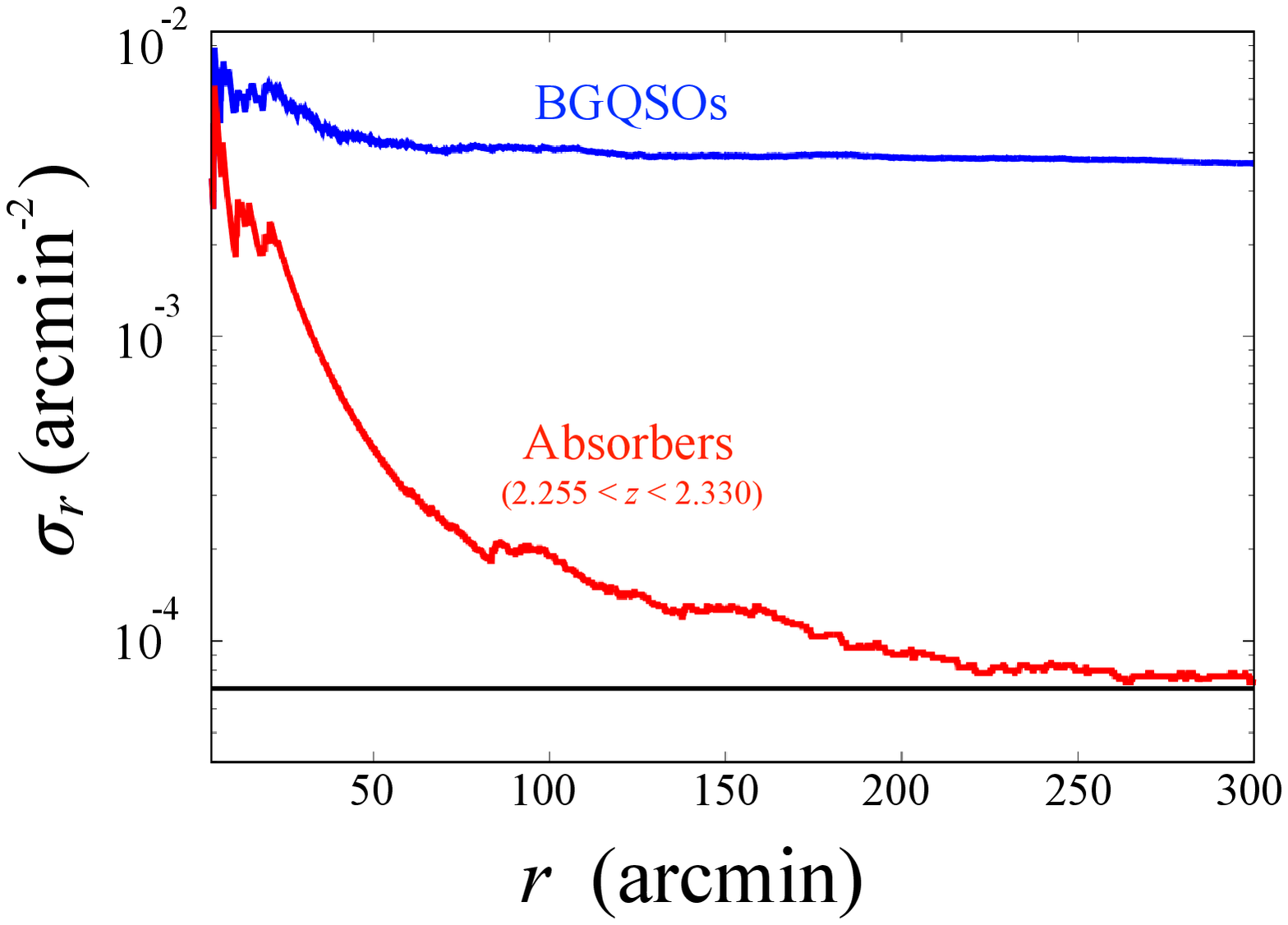} 
 \end{center}
\caption{The average surface number density of all 6 absorber concentrated regions, as a function of the radius ($r$).
The vertical axis indicates the surface number density of absorbers and BGQSOs
within a circular regions with a radius of $r$. 
The red and blue lines show the surface number density of absorbers and BGQSOs, respectively.
The black line indicates the average of absorber density calculated for the entire BOSS field.
}\label{fig:density}
\end{figure}

Figure \ref{fig:density} shows the average surface number density of absorbers and BGQSOs in all the 6 absorber-concentrated regions, 
as a function of the radius adopted to calculate the surface density. 
Here, we set the center of the circular region for the density calculation at the center of the position of absorbers.
For calculating the density, we simply divide the number of absorbers at $2.255 < z < 2.330$ by the area to calculate the surface number density. 
As shown in Figure \ref{fig:density}, the surface number density of absorbers rapidly decreases with increasing radius while that of BGQSOs 
appears to be much flatter.
The average surface number density of absorbers rapidly approaches to the average of surface number densities calculated for the entire 
BOSS field ($6.99 \times 10^{-5}$~arcmin$^{-2}$; the black line in Figure \ref{fig:density}).
The surface number density of absorbers exceeds 3 times the average absorber density for the whole BOSS field
at $r \sim 75^{\prime}$ (corresponding to $\sim$120~Mpc in the comoving scale at $z=2.3$)
and reach to the average density at $r\sim 250^{\prime}$ ($\sim$400~Mpc in the comoving scale at $z=2.3$).
Since the profile of the surface number density is systematically different between absorbers and BGQSOs,
we conclude that absorbers are intrinsically clustered, independent of the sky distribution of BGQSOs in 
each field of absorber concentrated regions.

\subsection{Observations and data reduction}
We carried out wide imaging observations for detecting LAEs in the J1230+34 field
on the 16th and 17th April 2016 (UT) using Suprime-Cam on the 8.2 m Subaru Telescope.
In our observations, we used both of the NB filter $NB400$ ($\lambda_{\rm eff} =4003$~\AA~and ${\rm FWHM} = 92$~\AA)
and $g^{\prime}$-band filter ($\lambda_{\rm eff} =4809$~\AA~and ${\rm FWHM} = 1163$~\AA; Miyazaki et al. 2002).
The NB400 filter can probe the Ly$\alpha$ emission of LAEs at $2.255 < z < 2.330$.
During our observing run, the typical seeing sizes (in FWHM) were $0^{\prime\prime}.7 - 1^{\prime\prime}.3$ 
for $NB400$ and $0^{\prime\prime}.6 - 1^{\prime\prime}.0$ for $g^{\prime}$-band.
The total on-source integration time for $NB400$ and $g^{\prime}$ are 4.6 hrs (300 sec exposure $\times$ 55 shots) 
and 1.0 hr (120 sec exposure $\times$ 30 shots), respectively.

\begin{figure}
 \begin{center}
  \includegraphics[width=8cm]{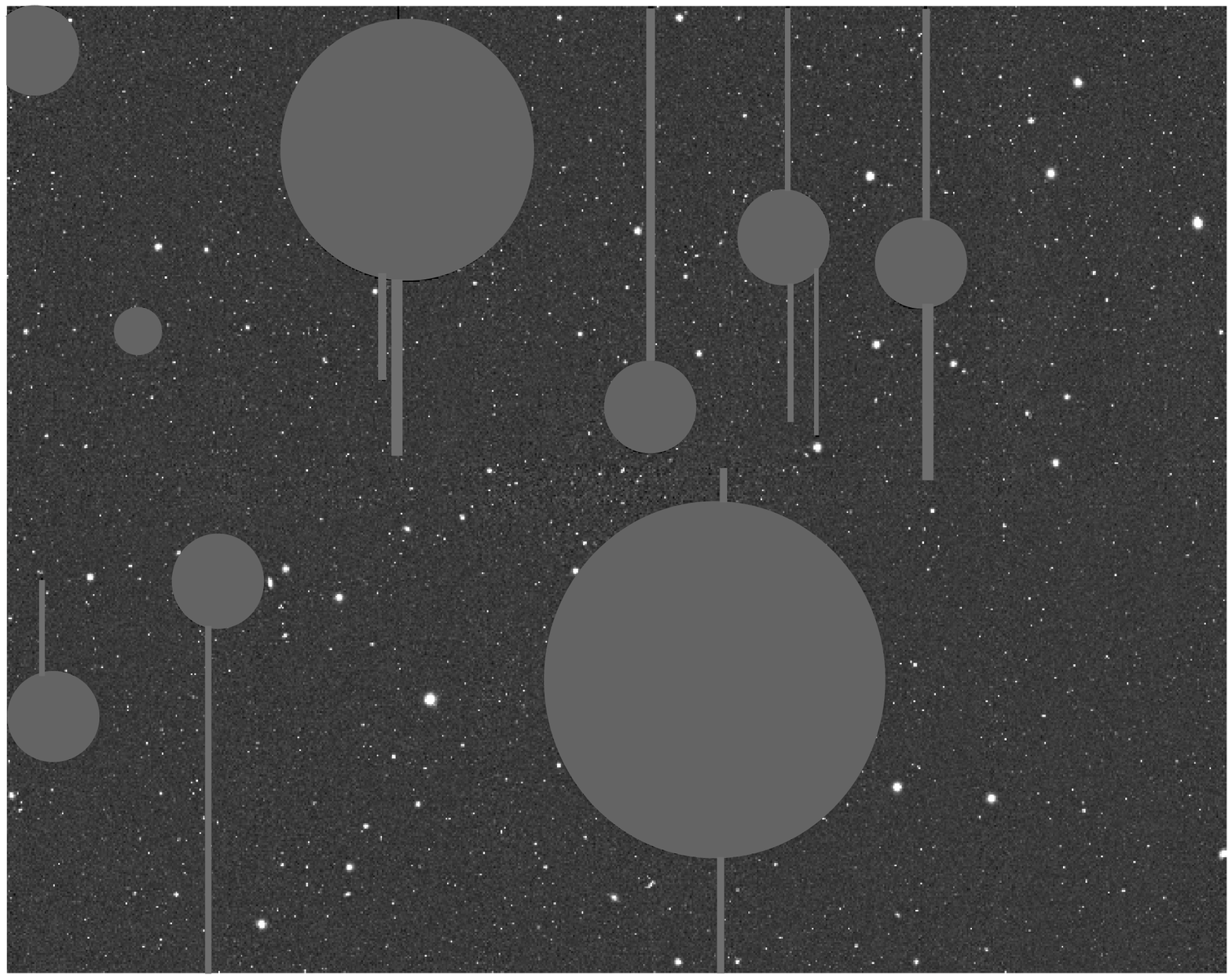} 
 \end{center}
\caption{
The reduced image of the J1230+34 field in NB400 (north is up). Gray filled circles and connected thin bars show masked regions.
}\label{fig:image}
\end{figure}

For the data reduction, we used SDFRED2 (Ouchi et al. 2004). 
The data reduction process with SDFRED2 is as follows. 
First, we subtracted bias and trimmed overscan regions from frames. 
Next, we applied the flat-fielding using self-flat frames\footnote{The flat frame made by stacking all the science frames 
without any shifts (i.e., astronomical objects should disappear in the finally created ``self-flat'' image).}
for both of $NB400$ and $g^{\prime}$ images. 
After the flat-fielding, we removed cosmic rays by using the L.A.Cosmic (van Dokkum 2001).
Corrections for the distortion and the atmospheric dispersion were then applied.
Based on the measurement of PSF sizes for point sources in each individual frame, we discarded some frames due to 
the bad seeing.
Consequently, we used 41 frames (corresponding to $\sim$3.4 hrs) and 28 frames ($\sim$0.9 hrs) of $NB400$ 
and $g^{\prime}$-band images for making the stacked images.
After subtracting the sky background, we made the mosaic images by calculating the median values.
After generating the mosaic images, we masked some regions affected by bright stars in either the $NB400$
or the $g^{\prime}$ images (see Figure \ref{fig:image}).

For the astrometry, we corrected R.A. and Dec. of objects in the J1230+34 field based on the USNO-B1 
catalog (Monet et al. 2003). 
After matching the position between the $NB400$ image and $g^\prime$ image, we smoothed the $g^{\prime}$-band 
image so that the PSF size of stars was matched to that of $NB400$ image.
The stellar PSF size of the smoothed image is $\sim0^{\prime\prime}.9$.
For the flux calibration, we used SDSS (DR12) spectroscopic stars within the observed field. We convolved transmission 
curves of $NB400$ and $g^{\prime}$ filters and spectra of 5 SDSS stars (Table \ref{tab:calib}) to estimate the photometric zero point.
We measured the photometric zero point based on synthetic magnitudes ($NB400_{\rm syn}$ and $g^{\prime}_{\rm syn}$ in Table \ref{tab:calib}) 
of those SDSS stars. For the calibration, we used the average of zero points measured using those 5 SDSS stars. 
We used SExtractor (Bertin \& Arnoults 1996) version 2.5.0 for the source detection and the photometry.
For the photometry, we used a circular aperture with a diameter of 2$^{\prime \prime}$, centered 
on the positions of the objects detected in the $NB400$ image.  
The 5$\sigma$ detection limit in the 2$^{\prime \prime}$ aperture for the $NB400$ and $g^{\prime}$ images 
are 25.34 and 26.79, respectively. 

We show the final image of $NB400$ in Figure \ref{fig:image}.
The field of view of the reduced image is $\sim 34^{\prime}.12 \times 26^{\prime}.33$, which is roughly
corresponding to $\sim 55~{\rm Mpc} \times 43~{\rm Mpc}$ in the comoving scale at $z=2.3$.
The area of the final image except for masked regions is $\sim 754$~arcmin$^{2}$.
The width of the NB400 wavelength coverage corresponds to $\sim$95~Mpc, and thus the surveyed 
volume is $\sim 1.9 \times 10^{5}$~Mpc$^{3}$ in the comoving scale at $z=2.3$.

\begin{table}
  \tbl{The SDSS stars used for the flux calibration}{%
  \begin{tabular}{llll}
      \hline
      Name & $g^{\prime}_{\rm model}$\footnotemark[a] & $g^{\prime}_{\rm syn}$\footnotemark[b] & $NB400_{\rm syn}$\footnotemark[b]\\ 
      \hline
      SDSS J123145.77+340537.5 & 20.44 & 20.36 & 21.91 \\
      SDSS J123136.24+341530.3 & 18.93 & 18.93 & 19.33 \\
      SDSS J123134.70+342057.4 & 18.92 & 18.90 & 19.31 \\
      SDSS J123138.71+340757.3 & 19.54 & 19.48 & 19.84 \\
      SDSS J122947.52+341933.6 & 18.97 & 18.95 & 19.43 \\
       \hline
    \end{tabular}}\label{tab:calib}
\begin{tabnote}
{\bf Notes.} \\
\footnotemark[a]$ g^{\prime}_{\rm model}$ is the model magnitude of SDSS. \\
\footnotemark[b]$ g^{\prime}_{\rm syn}$ and $NB400_{\rm syn}$ is the synthetic magnitudes determined by convoluting the filter response curves and spectrum of each star.
\end{tabnote}
\end{table}

\section{Results}

\subsection{Color selection of LAEs}
\begin{figure}
 \begin{center}
  \includegraphics[width=8.5cm]{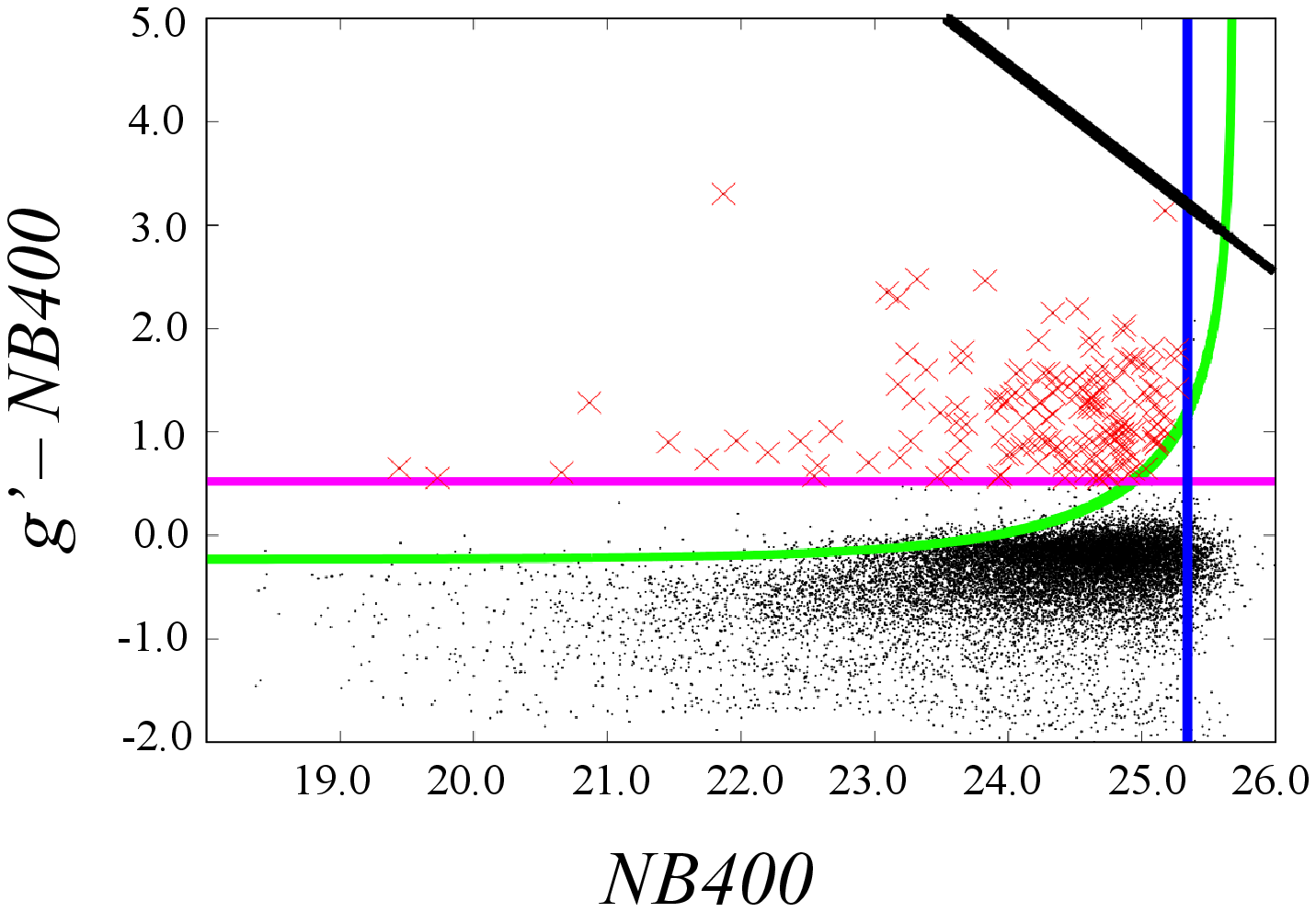} 
 \end{center}
\caption{$g^{\prime}-NB400$ vs. $NB400$ color-magnitude diagram. The $g^{\prime}-NB400$ color is based on
the aperture fixed (2$^{\prime\prime}$) photometry. The horizontal axis indicate $NB400$ magnitude from Kron photometry.
The colored lines indicate the selection criteria for LAEs. The magenta line indicates $g^{\prime}-NB400=0.522$ which corresponds to 
$EW_{0} = 20$ \AA. The green curve shows the 3$\sigma$ error in the $g^\prime - NB400$ color. The blue
line shows the 5$\sigma$ limiting magnitude of $NB400$. Red crosses denote the selected LAEs.
The black line corresponds to the 1$\sigma$ limiting magnitude of the $g^{\prime}$
image.
}\label{fig:selection}
\end{figure}

\begin{figure}
 \begin{center}
  \includegraphics[width=8.5cm]{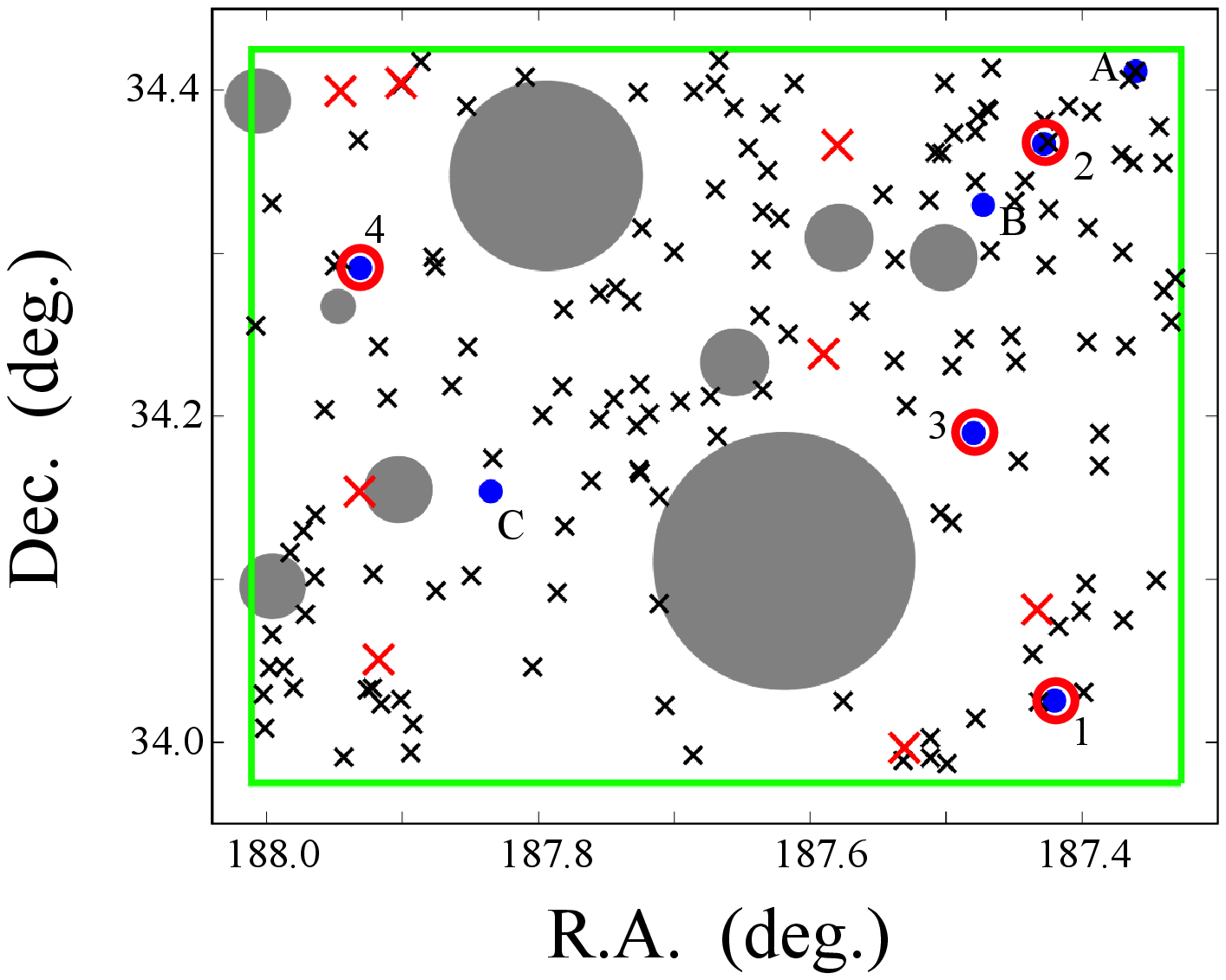} 
 \end{center}
\caption{
The sky distribution of BGQSOs (blue filled circles) and LAEs (black crosses) in the J1230+34 field. 
The green box shows the field-of-view of Suprime-Cam ($34^{\prime} \times 27^{\prime}$). 
The BGQSOs with surrounding red circles have absorbers at $2.255 < z < 2.330$.
LAEs shown by large red crosses have Ly$\alpha$ $EW_0 > 200$~\AA.
Masked regions are shown by gray filled circles.
IDs for absorbers and BGQSOs are same as those in Figure \ref{fig:map_j1230}.
}\label{fig:map_LAE}
\end{figure}

We select LAEs at $z=2.3$ based on the color-magnitude diagram, $g^{\prime}-NB400$ vs. $NB400$
(Figure \ref{fig:selection}). The adopted color criteria are as follows: (1) $g^{\prime}-NB400 \geqq 0.52$, 
(2) $NB400 \leqq 25.34$, and (3) $g^{\prime}-NB400 \geqq -0.23 + 3\sigma~{\rm in~color}$. 
The first criterion corresponds to the rest-frame equivalent width $EW_{0}$ of 20~\AA~at $z=2.3$.
The second  one is the 5$\sigma$ limiting magnitude for $NB400$.
The offset of $-0.23$ in the third criterion corresponds to the median color of the detected objects in the 
magnitude range of $24.0 < NB400 < 25.0$ except for the LAE candidates.
Then, we have selected 154 objects  as LAE candidates at $z=2.3$. After the visual inspection, 
the final sample consists of 149 LAE candidates. 
Since we are focusing on the very short wavelength range ($\sim 4000$~\AA) in optical, our sample is free from
the contamination by [O~{\sc iii}] emitters which could often be interlopers for high-$z$ LAEs.
The possible contaminants in our sample are [O~{\sc ii}]-emitting galaxies at $z \sim 0.07$, but such contamination is expected 
to be negligibly small based on the following considerations.
Given the 5$\sigma$ detection limit of the $NB400$ image and the NB-excess criterion for 
$EW_{0}$ ($g^{\prime}-NB400 \geqq 0.52$), the limiting [O~{\sc ii}] luminosity at $z\sim0.07$ is 
$\sim1.6\times 10^{38}$~erg~s$^{-1}$. 
The surveyed volume is 550 Mpc$^{3}$ in the comoving scale at $z\sim0.07$.
From the luminosity function (LF) of [O~{\sc ii}] emitters reported in Ciardullo et al. (2013), we obtain an expected 
number, $\sim$30 [O~{\sc ii}] emitters. Ciardullo et al. (2013) also reported that the distribution of $EW_{0}$ of 
[O~{\sc ii}] emitters at $z < 0.2$ shows its peak around 5~\AA, and rapidly decreases with the e-folding 
length of $w_{0} = 8$ \AA. We set the criterion of $EW_{0}$ for LAEs at $z \sim 2.3$ as $EW_{0} = 20$ \AA. 
This corresponds to $EW_{0} \sim 60~{\rm \AA}$  for [O~{\sc ii}] emitters at $z \sim 0.07$.
Based on this, the fraction of [O~{\sc ii}] emitters with $EW_{0} > 60~{\rm \AA}$ is only 0.055\% and the 
expected number of [O~{\sc ii}] emitters is $\sim0.018$. Thus we conclude that the 
contamination by [O~{\sc ii}] emitters is negligible, and we regard these 149 $NB400$-excess objects as 
LAEs at $2.255 < z < 2.330$ (see also Konno et al. 2016). We show their sky distribution  in Figure \ref{fig:map_LAE}.

\subsection{A candidate of a DLA counterpart}
\begin{figure}[h]
 \begin{center}
  \includegraphics[width=8cm]{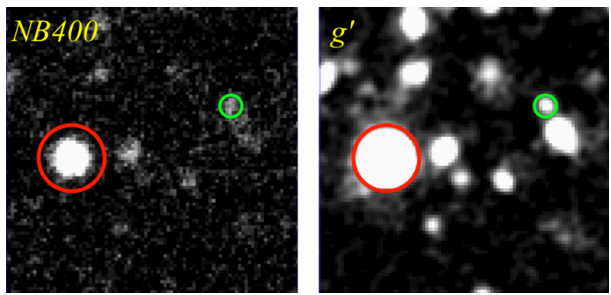} 
 \end{center}
\caption{Images of the candidate of the DLA counterpart in $NB400$ (left) and $g^{\prime}$ (right). The green and the red
circles show the candidate of the DLA counterpart and the corresponding BGQSO. The separation
of them is $10^{\prime\prime}.6$, that corresponds to $\sim$87~kpc at $z=2.3$.}
\label{fig:counterpart}
\end{figure}

So far, DLA counterparts have been mostly found close to quasar sight-lines with the impact parameter of 
$b \lesssim 25$~kpc (e.g., Fumagalli et al. 2010; Krogager et al. 2012; 2013; Kashikawa et al. 2014). 
On the other hand, Rao et al. (2011) investigate the distribution of
impact parameters at low-$z$ ($0.1 < z < 1$) and show that DLA counterparts could have impact parameters up to 
$\sim$100~kpc (see also Battistiet al. 2012; Straka et al. 2016). Motivated by this low-$z$ study, we search for counterparts of 
strong absorbers from our LAE sample within 
$12^{\prime\prime}.2$ (corresponding to 100 kpc at $z=2.3$) from the sight-line to showing a strong absorption feature.
As a result, we find a LAE near 1 DLA (along the line-of-sight to a quasar, the SDSS J122942.74+342202.1; No.2 in 
Table \ref{tab:info_j1230}, Figures \ref{fig:map_j1230}, and \ref{fig:map_LAE}) as shown in 
Figure \ref{fig:counterpart}, among 4 absorbers in the J1230+34 field. 
Its impact parameter to the quasar sight-line is $\sim$87~kpc (physical scale, that corresponds to $10^{\prime\prime}.6$). 
Note that, there are no objects with $g^{\prime}-NB400 > 0$ closer than $10^{\prime\prime}.6$ from the quasar sightline.
The apparent magnitudes of this candidate are $NB400 = 25.30$ and $g^{\prime} = 26.25$.
We derive the star-formation rate (SFR) of this candidate based on the Ly$\alpha$ flux 
($F_{\rm Ly\alpha}$ = 1.9 $\times$ 10$^{-17}$ erg s$^{-1}$ cm$^{-2}$, that corresponds to 
the Ly$\alpha$ luminosity of $L_{\rm Ly\alpha} = 7.7 \times 10^{41}$~erg~s$^{-1}$)
using the Kennicutt (1998) relation. 
The derived $SFR$ of 0.7~$M_{\odot}~{\rm yr}^{-1}$ is a bit lower than other detection of DLA counterparts at $z > 2$ 
reported by some previous studies (e.g., Fumagalli et al. 2010; P{\'e}roux et al. 2012; Bouch{\'e} et al. 2013; Fynbo et al. 2013; 
Krogager et al. 2013; Srianand et al. 2016).

To confirm whether or not this LAE is really a DLA counterpart, spectroscopic follow-up observations are required.
On the other hand, optical candidates of the counterparts are not found for the remaining 3 absorbers, at least as LAEs.
Further NB imaging data with deeper depth are required to explore the candidates for the counterparts of the remaining 3 absorbers.
Alternatively, counterparts for those absorbers may be other populations of galaxies such as Lyman break galaxies that cannot be 
selected through NB observations.
In this case, further $u$-band observations will be useful to identify galaxies through the so-called BMBX selection (Steidel et al. 2004)
without relying on the Ly$\alpha$ emission.

\subsection{The Ly$\alpha$ luminosity function}
Here, we derive the Ly$\alpha$ luminosity function (LF) of LAEs detected in the J1230+34 field and 
compare it with some previous results at a similar redshift.

\subsubsection{The Ly$\alpha$ flux and luminosity}
We calculate the Ly$\alpha$ flux and luminosity of the detected LAEs following the method described in 
Mawatari et al. (2012). The flux per unit frequency (${\rm erg~s^{-1}~cm^{-2}~Hz^{-1}}$) of the $NB400$ ($f_{\nu NB400}$) 
and the $g^{\prime}$-band (\textcolor{red}{$f_{\nu g^{\prime}}$}) are described as
\begin{equation}
  f_{\nu NB400} = f_{\nu {\rm C}} + f_{\nu {\rm Ly}\alpha}, \\
\end{equation}
\begin{equation}
  f_{\nu g^{\prime}} = f_{\nu {\rm C}} + f_{\nu {\rm Ly}\alpha} \times \frac{\int_{\lambda_{min,NB400}}^{\lambda_{max,NB400}} T_{\lambda,g^{\prime}}d\lambda} {\int_{\lambda_{min,g^{\prime}}}^{\lambda_{max,g^{\prime}}} T_{\lambda,g^{\prime}}d\lambda},
\end{equation}
where $f_{\nu {\rm Ly}\alpha}$ and $f_{\nu {\rm C}}$ are the flux per unit frequency of the Ly$\alpha$ and the 
continuum emission at $\lambda_{\rm obs}=4003$~\AA, respectively.
$T_{\lambda,g^{\prime}}$ is the response function of $g^{\prime}$-band filter.
Assuming the flat continuum over the range of the $g^{\prime}$-band coverage, we can calculate the Ly$\alpha$ flux and 
the $g^{\prime}$ flux. 
Based on the Ly$\alpha$ flux per unit frequency, we can calculate the integrated flux (${\rm erg~s^{-1}~cm^{-2}}$) as
\begin{equation}
  F_{\rm Ly\alpha} = f_{\nu {\rm Ly}\alpha} \times \Delta\nu_{NB400},
\end{equation}
where $\Delta\nu_{NB400}$ is the band width in FWHM (Hz) of the $NB400$ filter (1.72 $\times$ 10$^{13}$ Hz).
The Ly$\alpha$ luminosity is estimated as
\begin{equation}
  L_{{\rm Ly}\alpha} = 4\pi d_{\rm L}^{2}F_{{\rm Ly}\alpha},
\end{equation}
where $d_{\rm L}$ is the luminosity distance.
For calculating $d_{\rm L}$ of the LAEs in our study, we assume that all of LAEs are at $z=2.3$ that corresponds 
to the center of the $NB400$ coverage.

\subsubsection{Detection completeness}
\begin{figure}[h]
 \begin{center}
  \includegraphics[width=8cm]{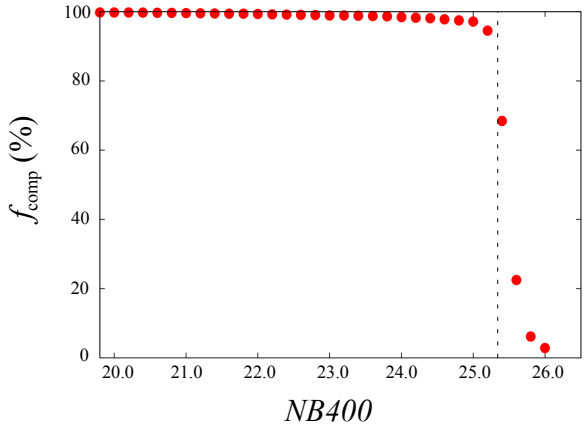} 
 \end{center}
\caption{The detection completeness, $f_{\rm comp}$, as a function of the $NB400$ magnitude.
The black dotted line shows the 5$\sigma$ limiting magnitude of $NB400$. 
}\label{fig:comp}
\end{figure}

We estimate the detection completeness for objects in the $NB400$ image using Monte Carlo simulations.
The procedure of estimation is as follows: (1) inserting artificial objects randomly in the $NB400$ image except for
masked regions, (2) detecting the artificially inserted objects with the same condition as the real detection with SExtractor 
(see Section 2.2), and (3) measuring the fraction of the number of successfully detected artificial objects as a function of 
the apparent $NB400$ magnitude from $NB400 = 19.8$ to 26.0 in 0.2 mag intervals.
Here we assume that distributed objects are point sources with FWHM of $0^{\prime\prime}.9$, that corresponds to the 
PSF size of the smoothed $NB400$ image (see Section 2.3).
Figure \ref{fig:comp} shows the detection completeness as a function of $NB400$ magnitude. 
Around the 5$\sigma$ limit of the $NB400$ magnitude, the estimated detection completeness is $\sim$70\%.
The completeness decreases quickly beyond the 5$\sigma$ limiting magnitude.

\subsubsection{The Ly$\alpha$ luminosity function}
We derive the Ly$\alpha$ LF of LAEs detected in our $NB400$ observation, based on the Ly$\alpha$ luminosity 
(Section 3.3.1) and detection completeness (Section 3.3.2). Note that we do not correct for the selection completeness 
(i.e., how LAEs are completely selected from the detected objects through the adopted color criteria), since 
Ouchi et al. (2008) showed that the incompleteness through the color selection is negligible.
We divide the completeness-corrected number of LAEs by the surveyed volume following the calculation method of previous NB studies
of Ly$\alpha$ LF of LAEs (e.g., Ajiki et al. 2003; Ouchi et al. 2003; 2008; Hu et al. 2004; Malhotra \& Rhoads 2004; Konno et al. 2014; 2016).
 
\begin{figure}
 \begin{center}
  \includegraphics[width=10cm]{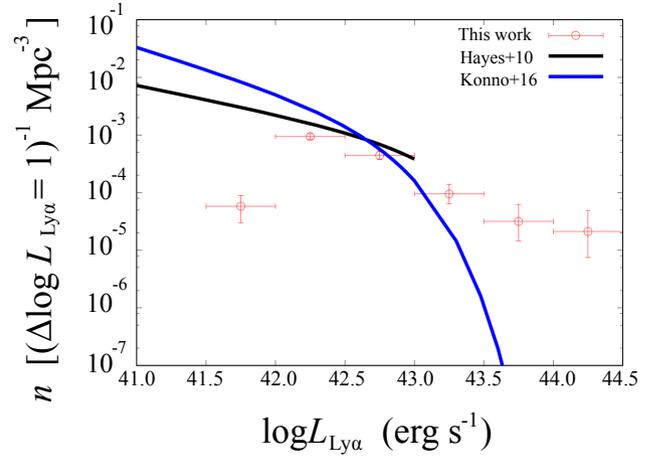} 
 \end{center}
\caption{The Ly$\alpha$ luminosity function. Red open circles show our result with a luminosity bin of $\Delta{\rm log}L_{\rm Ly\alpha}=0.5$. 
The blue and black curves show the Schechter function fits for LAEs at $z=2.2$ by Konno et al. (2016) and 
Hayes et al. (2010), respectively.}\label{fig:LF}
\end{figure}

Figure \ref{fig:LF} shows derived Ly$\alpha$ LF of our LAE sample. The Ly$\alpha$ LF of LAEs is usually fitted by the
Schechter function (Schechter 1976); i.e., the number density of LAEs tends to decrease exponentially with 
the Ly$\alpha$ luminosity at the bright side.
However, in the high luminosity regime ($L_{\rm Ly\alpha} > 10^{43.5}$ erg~s$^{-1}$) of our result, the number density 
of LAEs dose not show the exponential decrease but showing an excess of the number density with respect to the Schechter function.
This trend is seen also in some previous LAE studies (e.g., Ouchi et al. 2008; Konno et al. 2016), and it is thought to 
be due to the contribution of active galactic nuclei (AGNs). As for our case, we confirm that at least one object 
in the brightest luminosity bin is a quasar included in the BOSS quasar catalog (BGQSO-A, SDSS J122926.53+342441.9; 
P{\^ a}ris et al. 2012).
In the low-luminosity regime ($L_{\rm Ly\alpha} < 42.0$ erg s$^{-1}$), the correction for the completeness 
is not enough, due to very low selection completeness owing to relatively large photometric errors of faint objects. 

We then compare the derived Ly$\alpha$ LF of LAEs in the J1230+34 field with the Ly$\alpha$ LFs in blank fields.
Specifically, we compare our result with the Ly$\alpha$ LF of LAEs at $z=2.2$ reported by Konno et al. (2016) 
and Hayes et al. (2010) in Figure \ref{fig:LF}.
Due to the contribution of AGNs and the incompleteness we described above,  
we focus on the luminosity range of $10^{42.0} < L_{\rm Ly\alpha} < 10^{43.5}$ erg~s$^{-1}$ for the comparison.
Consequently, as shown in Figure \ref{fig:LF}, no significant difference in the Ly$\alpha$ LF of LAEs in the J1230+34 field 
relative to blank fields is found.

\subsection{The frequency distribution of Ly$\alpha$ $EW_{0}$}
\begin{figure}
 \begin{center}
  \includegraphics[width=8cm]{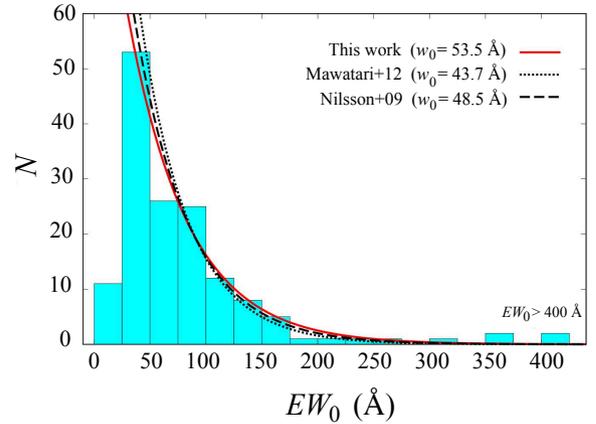} 
 \end{center}
\caption{The frequency distribution of the Ly$\alpha$ rest-frame equivalent width. The bin width for $EW_{0}$ is 
25 \AA. The red solid, black dotted, and black dashed lines show the exponential fit for this work, Mawatari et al. (2012)
at $z=2.4$, and Nilsson et al. (2009) at $z=2.25$, respectively.
}\label{fig:hist_EW}
\end{figure}

In order to characterize properties of LAEs in the J1230+34 field, we study their $EW_{0}$ distribution (Figure \ref{fig:hist_EW}). 
Here, we compare our result with that of LAEs at a similar redshift ($z=2.4$) in an over-density region
of LAEs reported by Mawatari et al. (2012) and in a blank field at $z=2.25$ reported by Nilsson et al. (2009).
For a quantitative comparison of the Ly$\alpha$ $EW_0$, we adopt the following analytic formula,
\begin{equation}
N = C \times e^{-EW_{0}/\omega_{0}}
\end{equation}
(see Gronwall et al. 2007; Mawatari et al. 2012).
We fit the Ly$\alpha$ $EW_{0}$ distribution in the $EW_{0}$ range of $25 \leq EW_{0} \leq 350$ \AA.
The $e$-folding lengths, $w_{0}$, of this work, Mawatari et al. (2012), and Nilsson et al. (2009) are 
${w_0}=53.5 \pm 4.7$~\AA,  ${w_0}=43.7 \pm 0.4$~\AA, and ${w_0}=48.5 \pm 1.7$~\AA, respectively. 
We can mention that $w_{0}$ in our study is consistent with those in Mawatari et al. (2012) and
Nilsson et al. (2009) at 2$\sigma$ and 1$\sigma$, respectively.

\subsection{LAEs with a large $EW_{0}$}
In this work, we are focusing on LAEs which are thought to be young population of galaxies.
This is because the absorber-concentrated regions are possibly gas-rich regions and thus young galaxies could exist there. 
Since the younger LAEs tend to show the larger Ly$\alpha$ $EW_{0}$ (e.g., Malhotra \& Rhoads 2002; Nagao et al. 2007), 
we here focus on LAEs with a large $EW_{0}$.
In Table \ref{tab:largeEW}, we show the basic data of LAEs with $EW_{0} > 200$~\AA.
There are 8 LAEs with $EW_{0} > 200$~\AA~in the J1230+34 field. All of their positions are somewhat far from absorbers.
The nearest LAE with large $EW_{0}$ is 5.5 Mpc (corresponding to $\sim203^{\prime\prime}$) away from an absorber. 
There are no large $EW_{0}$ LAEs within 5 Mpc (corresponding to $\sim185^{\prime\prime}$) from each absorber.

{\footnotesize
\begin{table*}
  \tbl{General information of LAEs with $EW_{0} > 200$~\AA}{%
  \begin{tabular}{cccccccccccc}
      \hline
R.A. & Dec. & $NB400$\footnotemark[a] & $NB400$\footnotemark[b] & $g^{\prime}$\footnotemark[a] & $g^{\prime}$\footnotemark[b] & $EW_0$ & log$L_{\rm Ly\alpha}$\footnotemark[c]  & $d_{\rm abs-LAE}$\footnotemark[d] & $d_{\rm abs-LAE}$\footnotemark[d] & Nearest\footnotemark[e] \\
(deg.) & (deg.) & (ap) & (AUTO) & (ap) & (AUTO) & (\AA) & (erg~s$^{-1}$) & (arcmin) & (Mpc) & absorber \\
      \hline
187.9011 & 34.4046 & 23.08& 21.86 & 26.39 & 24.11 & 1381.4 & 43.61 & 6.91 & 11.2 & 4 \\
187.9315 & 34.1540 & 25.37 & 25.17 & $>$27.35\footnotemark[1] & -\footnotemark[2] & $>$1002.7\footnotemark[3] & 42.34$<$\footnotemark[4]& 8.15 & 13.2 & 4 \\
187.5902 & 34.2382 & 23.54 & 23.32 & 26.01 & 25.84 & 365.3 & 43.04 & 6.14 & 10.0 & 3 \\
187.5307 & 33.9967 & 24.88 & 23.83 & 27.35 & 25.37 & 357.7 & 42.77 & 5.71 & 9.3 & 3 \\
187.4330 & 34.0814 & 23.20 & 23.10 & 25.55 & 25.37 & 308.1 & 43.12 & 3.37 & 5.5 & 1 \\
187.9175 & 34.0508 & 24.58 & 24.52 & 26.77 & 26.34 & 250.8 & 42.52 & 14.29 & 23.2 & 3 \\
187.9455 & 34.3996 & 24.65 & 24.33 & 26.80 & 25.72 & 238.0 & 42.54 & 6.49 & 10.5 & 4 \\
187.5800 & 34.3663 & 25.06 & 24.87 & 27.09 & 26.91 & 203.5 & 42.40 & 7.45 & 12.1 & 2 \\
     \hline
    \end{tabular}}\label{tab:largeEW}
\begin{tabnote}
 {\bf Notes.} \\
\footnotemark[a] $NB400$ and $g^{\prime}$ magnitudes measured by the fixed-aperture (2$^{\prime\prime}$) photometry. \\
\footnotemark[b] $NB400$ and $g^{\prime}$ magnitudes measured by the Kron photometry. \\
\footnotemark[c] The Ly$\alpha$ luminosity based on the Kron photometry. \\
\footnotemark[d] The distance from the LAE to the nearest absorber. \\
\footnotemark[e] The ID of the nearest absorber from the LAE. The number corresponds to the No. in Table \ref{tab:info_j1230}, Figures \ref{fig:map_j1230}
and \ref{fig:map_LAE}. \\
\footnotemark[1] The 3$\sigma$ limiting magnitude. \\
\footnotemark[2] The measured flux in the $g^{\prime}$-band of this object is negative. \\
\footnotemark[3] The lower limit of $EW_0$ calculated from the 3$\sigma$ limiting fixed-aperture magnitude in the $g^{\prime}$-band.\\
\footnotemark[4] The upper limit of $L_{\rm Ly\alpha}$ calculated by assuming $f_{\nu}=0$ in the $g^{\prime}$-band.
\end{tabnote}
\end{table*}
}

\begin{figure}
 \begin{center}
  \includegraphics[width=8cm]{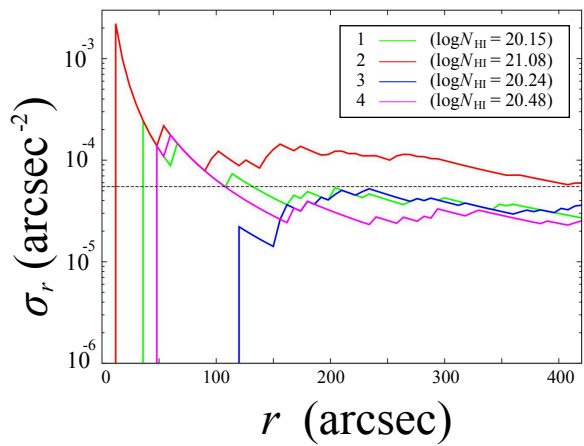} 
 \end{center}
\caption{The surface number density of LAEs around each absorber in the J1230+34 field, as a function of the radius ($r$) 
adopted to calculate the surface density. 
The vertical axis indicates the surface number density of LAEs within a circular regions at  $r$. 
The horizontal axis indicates the radius of circular regions from the position of each absorber. 
The IDs for the absorbers correspond to those in Table \ref{tab:info_j1230} and Figure \ref{fig:map_j1230}.
The dotted horizontal black line indicates the average of the LAE density calculated for the entire J1230+34 field.
}\label{fig:LAE_density}
\end{figure}

\begin{figure}
 \begin{center}
  \includegraphics[width=8cm]{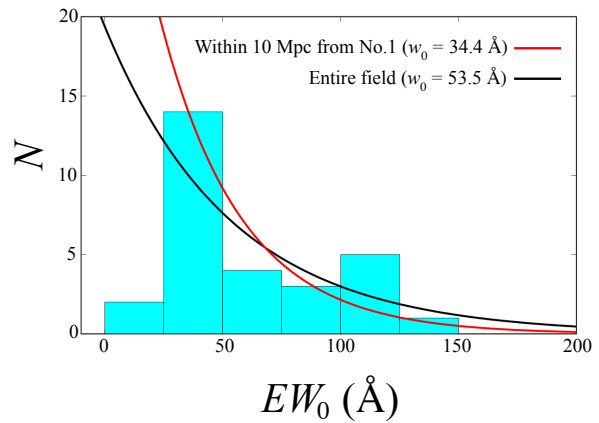} 
 \end{center}
\caption{The frequency distribution of the Ly$\alpha$ $EW_{0}$ of LAEs 
within 10 Mpc from the absorber No. 2 (red line) and in the entire target field (black line).
}\label{fig:EW0_10Mpc}
\end{figure}

\section{Discussion}

\subsection{The properties of LAEs located close to absorbers}
As described in Sections 3.3 and 3.4, the obtained Ly$\alpha$ LF is consistent with those in blank fields at 
similar redshift. Also, we found no differences in the frequency distribution of the Ly$\alpha$ $EW_{0}$ in the target
field and those in some other fields including an overdensity region and blank fields.
Here we focus on the results of Mawatari et al. (2012). 
They studied the spatial distribution of LAEs around a radio galaxy at $z\sim2.4$ (53W002) using Subaru/Suprime-Cam.
They found that there is a significantly high density region around the radio galaxy spreading over 
$\sim 5^{\prime} \times 4^{\prime}$ (8.3 Mpc $\times$ 6.6 Mpc in the comoving scale at $z=2.4$). 
However, the number density of LAEs in the entire the 53W002 field ($\sim$50 Mpc $\times$ 40 Mpc) 
is comparable to the average density of LAEs in blank fields at $z\sim2$. 
This implies that we may miss the density excess in such Mpc-scale region by focusing only on the average density in the entire field.
Therefore, it is worthwhile to study local characteristics of LAEs in the J1230+34 field. 
Motivated by this, we investigate the surface number density of LAEs around absorbers in the target field.
Figure \ref{fig:LAE_density} shows the surface number density of LAEs around each absorber in the J1230+34 field.
The surface number density of LAEs is calculated by dividing the number of LAEs by area of circular region centered on the
positions of each absorber.
As seen in Figure \ref{fig:LAE_density}, LAEs around the absorber No.2 show the density excess over $\sim$400 arcsec ($\sim$11 Mpc) 
relative to the average density of LAEs in the entire J1230+34 field (black dotted line in Figure \ref{fig:LAE_density}).
Indeed, in Figure \ref{fig:map_LAE}, we can see the possible number excess of LAEs around the absorber No. 2.
Interestingly, the absorber No. 2 is the strongest DLA with $N_{\rm HI}$ (log$N_{\rm HI} = 21.08$~cm$^{-2}$) in the J1230+34 field.
On the other hand, we found no density excess of LAEs around remaining 3 absorbers, whose H~{\sc i} column densities are
smaller than that of absorber No. 2.
Here we should mention the BGQSOs around the absorber No.2.
The BGQSO A (in Figure \ref{fig:map_LAE}) is a quasar at $z=2.275$ and thus it is a member of objects in the absorber-concentrated region. 
Indeed, this BGQSO A is detected as a LAE in our observation (see Section 3.3.3). 
The projected comoving distance from the BGQSO B to the absorber No.2 is  $\sim$6.9 Mpc (255$^{\prime\prime}$) and the redshift 
difference ($\Delta z$=0.007) between the BGQSO A and the absorber No2 corresponds to $\sim$9.0 comoving Mpc. 
Since we set the redshift range of the BGQSOs not to miss PDLAs, this quasar satisfy the criteria even its redshift is very 
similar with those of target absorbers.
As for the BGQSO B, although it satisfies the BGQSO criteria, it is difficult to recognize intervening absorbers at
$z\sim2.3$ due to low S/N since its Lyman limit locates very close to the wavelength coverage of $NB400$ as shown in Figure \ref{fig:spBGQSO}.
Except for BGQSO-A and B, 4 among 5 BQSOs in the J1230+34 field have strong Ly$\alpha$ absorber at $2.255 < z < 2.330$. 

We also examine the frequency distribution of the Ly$\alpha$ $EW_{0}$ within 10 Mpc from the absorber No. 2.
There are 29 LAEs within 10 Mpc from the absorber No. 2.
Figure \ref{fig:EW0_10Mpc} shows the derived frequency distribution of the Ly$\alpha$ $EW_0$ around the absorber No. 2. 
By adopting the same fitting formula as in Section 3.4 (Equation 5), we obtain the exponential length of $w_0 = 34.4 \pm 10.1$.
Although the value of $w_{0}$ is smaller than that in the entire target field ($w_0 = 53.5 \pm 4.7$), $w_{0}$ around absorber No.2 is 
consistent with that in the entire field within 2$\sigma$ due to the large error. Here, the fitted $EW_{0}$ range is $25 \leq EW_{0} \leq 200$ \AA. 
Out of 29 LAEs within 10 Mpc from absorber No. 2,  $\sim$80\% of them have $EW_{0}<100$~\AA. 
There are no LAEs with $EW_{0}>150$~\AA~close to the DLA (see also Table \ref{tab:largeEW}). 
There are no difference in the properties of LAEs in the large scale, while we find a possible overdenstiy of LAEs 
in the small scale of $\sim$10 Mpc. 
Based on this finding, we discuss possible scenarios for absorber-concentrated regions in Section 4.3.

\subsection{A quasar in the possible overdensity region of LAEs around the absorber No.2}
As we report in Section 4.1, we find a quasar associate with the possible overdensity of LAEs.
It has been thought that the quasar is a good tracer for high-$z$ oversnsity regions (e.g., Wylezalek et al. 2013; Adams et al. 2015).
However, some recent studies (e.g., Ba{\~n}ados et al. 2013; Mazzucchelli et al. 2016; Uchiyama et al. 2017) show that many 
quasars do not reside in overdensity environment.
Especially, Uchiyama et al. (2017) studied environments of $>$150 quasars using the wide-field ($>$100 deg$^2$) data
obtained with Hyper Suprime-Cam (HSC; Miyazaki et al. 2012) on the Subaru telescope.
They reported that most quasars in their sample do not reside in overdensity environments but in general environments.
These recent works imply that the possible LAE overdensity we found may rather associate with absorber environment
and may not be related to the quasar environment.

\subsection{Possible scenarios for absorber-concentrated regions}
Here, we discuss why the Ly$\alpha$ LF and the Ly$\alpha$ $EW_{0}$ distributions in the J1230+34 field show no difference
compared to those in an overdensity region and blank fields.
One possible scenario is that the J1230+34 field harbors many young galaxies intrinsically, that are made from plenty of gas 
suggested by the concentration of strong absorbers.
However, since there is a large amount of H~{\sc i} gas around absorbers, the effect of the resonant scattering is serious.
Since the Ly$\alpha$ emission is the resonant line and has a large scattering cross section, the optical depth becomes
very high even with small amount of gas (e.g., Hayes et al. 2015). Therefore, Ly$\alpha$ photons experience a large number
of scattering, resulting in a higher probability that Ly$\alpha$ photons are absorbed by dust.
Consequently, we may underestimate the Ly$\alpha$ $EW_{0}$ and miss a large fraction of LAEs around absorbers.
On the other hand, as we show in Section 4.1, we find a possible overdensity of LAEs around a DLA with the highest $N_{\rm HI}$
in the target field while no overdensity is found around absorbers with smaller $N_{\rm HI}$. 
At a glance, this situation looks inconsistent with the scenario described above. 
However, we may explain both of the existence of LAE overdensity around the DLA with the highest $N_{\rm HI}$
and the lack of the number excess of LAEs around absorbers with lower $N_{\rm HI}$ as follows:
(1) around high $N_{\rm HI}$ absorber, there are enough number of LAEs to detect the overdensity even though the effect of
the resonant scattering is serious, 
and (2) around lower $N_{\rm HI}$, although there are intrinsic number excess of LAEs, it is not enough to defeat the effect of 
resonant scattering.
To examine this scenario, further H$\alpha$ observations are required since it is basically free from the resonant 
scattering (e.g., Garn \& Best 2010). If this scenario is the case, we will find the number excess of H$\alpha$ emitters (HAEs)
around absorbers.

An alternative idea is that there are not so many galaxies in the target field even though there are many neutral gas.
H~{\sc i} gas is not converted stars directly but goes through the molecular clouds. Thus, a large fraction of H~{\sc i}
gas may not be converted to stars. To discuss the star-formation process in the absorber-concentrated region,
observations of molecular gas are required as well as H~{\sc i} gas and galaxies.

Furthermore, if we follow the idea that the origin of absorbers is the disk of galaxies, gas may not extend so widely.
In this case, an absorber corresponds just to a galaxy and thus the absorber-concentrated region may not be
a gas-rich region. Although the fact that BGQSOs (A and B) near absorber No.2 do not have absorbers at $z\sim2.3$
may support this idea, BGQSO-A is at $z=2.275$ and thought to be a member of $z\sim2.3$ objects and we
cannot recognize absorbers $z\sim2.3$ on the spectra of BGQSO-B. Therefore, we cannot discuss this possibility
only with our current data.
To examine this idea, the surface number density of the BGQSOs is still low.
When the data of future deep spectroscopic surveys such as the extend BOSS (eBOSS; Dawson et al. 2015) and the 
Subaru Prime Focus Spectrograph (PFS) survey (Takada et al. 2014) will be available, the surface number density 
of BGQSOs will increase significantly and accordingly we can tackle this problem.
Alternatively, DLAs whose background sources are galaxies (gal-DLAs) are newly found by Cooke \& O'Meara (2015) and  
Mawatari et al. (2016). Since the surface density of galaxies is much higher than that of quasars, we can investigate
the spatial distribution of neutral gas by focusing on gal-DLAs. Future 30-m class telescopes enable us to solve this problem.

\section{Concluding remarks}
Based on the BOSS strong Ly$\alpha$ absorber catalog, we have searched for absorber-concentrated regions and obtained the following results.

\begin{enumerate}
\item[1.] We find 6 concentrated region of strong Ly$\alpha$ absorbers $N_{\rm H~I} > 10^{20.0} {\rm cm}^{-2}$ from the absorber 
catalog based on the BOSS, i.e., three or more absorbers distribute within the cubic region of (50~Mpc)$^3$ with the comoving scale 
in the redshift range of $2.255 < z < 2.330$.
\item[2.] Among 6 absorber concentrated regions, we found a rare and interesting region, J1230+34 field, where 4 absorbers 
(2 DLAs and 2 sub-DLAs) distribute within the cuboidal space with 50 Mpc $\times$ 50 Mpc $\times$ 53.5 Mpc. 
\item[3.] Through a wide and deep Suprime-Cam search for LAEs with the $NB400$ filter toward the J1230+34 field, 
149 LAEs are detected.
\item[4.] We find no difference in the Ly$\alpha$ LF in the target field relative to those in blank fields at similar redshift.
\item[5.] The frequency distribution of the Ly$\alpha$ $EW_0$ of the LAEs in the entire J1230+34 field is 
similar to those in a overdensity region and blank field at $z\sim 2$ but much smaller than those in higher redshift.
\item[6.] When we focus on the region close to absorber, we find a possible number excess of LAEs around
a DLA with the highest H~{\sc i} column density (log$N_{\rm HI} = 21.08$~cm$^{-2}$) in the target field.
\end{enumerate}

The target field, the J1230+34 field is an important region to understand the early phase of galaxy evolution.
Further observations in the different wavelength range, such as H$\alpha$ emission, are expected to investigate
the properties of galaxies around absorbers.
Unfortunately, it is still challenging to investigate the extent of the H~{\sc i} gas due to too low number density
of BGQSO even by using BOSS data. When the surface number density of background sources increase
(not only quasars but also galaxies), the J1230+34 field is one of the most interesting field to investigate the 
relation of gas and galaxies.

\section*{Acknowledgments}
We thank to the anonymous referee for many fruitful comments and suggestions.
This paper is based on observations with the Subaru Telescope operated by National Astronomical Observatory 
of Japan (NAOJ).  We thank the Subaru staffs for their support during the observations. KO was supported by 
grants from JSPS (16J02689), NAOJ Visiting Fellow Program, and the Hayakawa Satio Fund awarded by the 
Astronomical Society of Japan.
TN is supported by grants from JSPS (KAKENHI grant Nos. 25707010, 16H01101, and 16H03958)  and by 
the JGC-S Scholarship Foundation. MI, KN and YT are also supported by JSPS KAKENHI
(grant Nos. 15K05030, 15H03645, and 23244031 and 16H02166, respectively).
We thank Ichi Tanaka at NAOJ, Ken Mawatari at Osaka Sangyo University, and Yasuhiro Shioya for useful 
comments and discussion.
Funding for SDSS-III has been provided by the Alfred P. Sloan Foundation, the Participating Institutions, 
the National Science Foundation, and the U.S. Department of Energy Office of Science. 
The SDSS-III web site is http://www.sdss3.org/.
SDSS-III is managed by the Astrophysical Research Consortium for the Participating Institutions of the 
SDSS-III Collaboration including the University of Arizona, the Brazilian Participation Group, Brookhaven 
National Laboratory, Carnegie Mellon University, University of Florida, the French Participation Group, 
the German Participation Group, Harvard University, the Instituto de Astrofisica de Canarias, the Michigan 
State/Notre Dame/JINA Participation Group, Johns Hopkins University, Lawrence Berkeley National Laboratory, 
Max Planck Institute for Astrophysics, Max Planck Institute for Extraterrestrial Physics, New Mexico State 
University, New York University, Ohio State University, Pennsylvania State University, University of Portsmouth,
Princeton University, the Spanish Participation Group, University of Tokyo, University of Utah, Vanderbilt University,
University of Virginia, University of Washington, and Yale University. 


\end{document}